\documentclass[aps,prc,twocolumn,superscriptaddress,floatfix,amsmath,amssymb,nofootinbib,longbibliography]{revtex4-2}
\usepackage[utf8]{inputenc}
\usepackage{amsmath}
\usepackage{graphicx}
\usepackage{lipsum}
\usepackage{listings}
\usepackage{listingsutf8}
\usepackage{xcolor}
\usepackage{amssymb}
\usepackage{textcomp}
\usepackage{units}
\usepackage{bbm}
\usepackage{enumitem}  
\usepackage{url}
\usepackage[english]{isodate}
\usepackage{bm}
\usepackage{mdframed}
\usepackage{dsfont}
\usepackage{dcolumn}
\usepackage{physics}
\usepackage{color}
\usepackage{placeins}
\usepackage{slashed}
\usepackage[caption=false]{subfig}
\usepackage{braket}
\usepackage{hyperref}
\usepackage[capitalise]{cleveref}
\usepackage{orcidlink}
\usepackage{hyperref}
\hypersetup{
    colorlinks = true,
    citecolor  = blue,
    linkcolor  = blue,
    urlcolor   = blue
}
\usepackage{mathrsfs}
\usepackage[normalem]{ulem}

\newcommand{\SigAnom}{ \Sigma^{12(\infty)} }
\newcommand{\SpVars}{\mathbf{p}m_tm_s}
\newcommand{\SpMomIso}{\mathbf{p}m_t}
\newcommand{\SpIso}{m_t}

\begin{document}

\title{Modeling quasielastic lepton-nucleus interactions with \emph{ab initio} spectral functions from infinite nuclear matter}
\author{Alma L. Cavallin \orcidlink{0009-0005-4110-6287}} 
\email{alma.cavallin@chalmers.se}
\affiliation{Department of Physics and Astronomy, Chalmers University of Technology, SE-412 96, Göteborg, Sweden}

\author{Francesco Marino \orcidlink{0000-0001-7743-1982}}
\affiliation{Institut f\"{u}r Kernphysik and PRISMA+ Cluster of Excellence, Johannes Gutenberg-Universit\"{a}t Mainz, 55128 Mainz, Germany}

\author{Joanna E. Sobczyk\orcidlink{0000-0003-4698-9339}}
\affiliation{Department of Physics and Astronomy, Chalmers University of Technology, SE-412 96, Göteborg, Sweden}

\date{24th August 2026}
\noindent
\begin{abstract} 
We present a study of quasielastic lepton-nucleus scattering within the local density approximation, using \emph{ab initio} spectral functions from infinite nuclear matter derived with self-consistent Green's functions theory and interpolated by neural networks. We include final-state interactions in terms of the particle spectral function and investigate the importance and range of validity of this treatment for varying momentum transfer. The performance of the model is tested for inclusive electron scattering on different isospin-symmetric target nuclei, and we present results for $^{12}$C, $^{16}$O, and $^{40}$Ca. Theoretical uncertainties originating from the neural network interpolation and the Hamiltonian dependence are assessed. We also present a calculation of charged-current responses of $^{16}$O and the flux-averaged total $\nu_\mu-^{12}$C cross section. Our model shows good agreement with experimental data within the range of validity of our approximations. The framework can be readily extended to include additional dynamical mechanisms, such as pion production, as well as other nuclear Hamiltonians. 

\end{abstract}

\maketitle

\section{Introduction}
The goals of next-generation neutrino oscillation experiments, such as the Deep Underground Neutrino Experiment (DUNE)~\cite{DUNE:2020lwj}  and the Tokai-to-Hyper-Kamiokande (T2HK) project~\cite{Hyper-Kamiokande:2018ofw}, include the determination of the mass hierarchy and of the charge-parity violating phase. The success of these high-statistics measurements depends on simultaneously reducing and controlling the systematic uncertainties arising from the theoretical modeling of neutrino-nucleus interactions~\cite{NuSTEC:2017hzk, Ankowski:2016jdd}. 

Two central challenges are the consistent inclusion of the different interaction channels within a single framework and the treatment of final-state interactions (FSI). The former arises because neutrino beams are not monoenergetic, so that a variety of reaction mechanisms contribute to the relevant neutrino-nucleus observables~\cite{NuSTEC:2017hzk}. The latter is tied to the design of neutrino experiments, which infer the neutrino energy from the distributions of the outgoing hadrons and of the charged lepton.
Since a fully quantum-mechanical calculation of semi-exclusive cross sections is at present generally not feasible (see, e.g.,~\cite{Bacca:2014tla,Roggero:2019myu,Rocco:2020jlx}), the Monte Carlo event generators used by experimental collaborations~\cite{Andreopoulos:2009rq,Hayato:2009zz,Golan:2012rfa,Buss:2011mx} model the scattering process in two steps: the weak interaction at the primary vertex, followed by hadron rescattering treated within the local density approximation (LDA).

Various theoretical approaches have been proposed and subsequently implemented in Monte Carlo generators. Among them, the spectral function (SF) formalism has been widely used, with the hole SF obtained either from microscopical calculations or from experimental data~\cite{Benhar:1994hw, Benhar:2006wy, Nieves:2017lij, Barbieri:2019ual,CLAS:2022odn, Sobczyk:2022ezo}. 
The SF formalism extends naturally to the relativistic regime and can accommodate higher-energy mechanisms beyond quasielastic (QE) scattering within the same framework~\cite{Rocco:2018mwt}. In many cases, however, FSI are either neglected or included only approximately, e.g., through an optical potential fitted to data \cite{Horikawa:1980, Ankowski:2014yfa}.

While many frameworks give fairly good descriptions of the QE mechanism, they struggle to obtain the same level of agreement with experimental data at the dip and $\Delta$ regions, and at higher energies. In many cases, the pion-production mechanism is still described in a simplified way, neglecting any in-medium treatment of baryon resonances and pions. To date, the only microscopic calculation of electroweak pion production in nuclei that includes an in-medium treatment of the $\Delta$ resonance is the LDA-based model of the Valencia group~\cite{Oset:1987re, Hernandez:2013jka}. The same model is widely used to describe pion rescattering in Monte Carlo generators.

Motivated by this landscape, we propose a new approach in which SFs for infinite symmetric nuclear matter (NM) are derived \emph{ab initio} within self-consistent Green's function (SCGF) theory (see~\cite{Rios:2020oad,Barbieri:2016uib}) using the Gorkov framework of Refs.~\cite{Marino:2024tfp,FrancescoMarino:2026dzw} with realistic nuclear Hamiltonians from chiral effective field theory~\cite{Ekstrom:2015rta,Jiang:2020the,Machleidt:2024bwl,Epelbaum2024}.
So far, \emph{ab initio} calculations of electroweak cross sections were performed only for inclusive observables using both the Green's function Monte Carlo method~\cite{Lovato:2014eva, Lovato:2015qka,  Lovato:2020kba} and recently the coupled-cluster theory~\cite{Sobczyk:2021dwm, Acharya:2024xah}. In a complementary line of work, the hole SFs for several nuclei of interest were obtained from various \emph{ab initio} methods~\cite{Rocco:2018mwt,Barbieri:2019ual,Sobczyk:2022ezo,Sobczyk:2023mey}. 

Here, we adopt the LDA and include not only the hole SF, which describes the initial state, but also the particle SF, through which FSI are accounted for. This allows for a more consistent description in the low- and medium-momentum transfer regime and constitutes a considerable improvement upon the available calculations employing \emph{ab initio} hole SFs. Although the LDA sacrifices the detailed shell structure of nuclei, it is well suited to neutrino oscillation experiments: the incoming neutrino flux is broad in energy, so that nuclear structure details are effectively smeared out. The LDA further allows us to treat isospin-symmetric nuclei, such as $^{12}$C and $^{16}$O, within the same theoretical framework, which is particularly relevant for the T2K and future T2HK experiments, whose near and far detectors use these nuclei as targets. Moreover, our LDA-based framework allows for a natural extension to include the in-medium pion production from the calculations already available. It can also be readily combined with intranuclear-cascade models of hadron rescattering, since these rest on the same underlying assumptions.

In this work, we set up our framework and benchmark it in the QE regime for both the electron- and neutrino-scattering data. We develop artificial neural networks working effectively as interpolators for the \emph{ab initio} SF data.
In \cref{sec:formalism}, the theoretical formalism is introduced, including the employed approximations and the treatment of FSI. We then describe the neural network training data, which are obtained from Ref.~\cite{FrancescoMarino:2026dzw}, in~\cref{sec:data}. Next, in~\cref{sec:ANN} we outline the construction of the neural networks and evaluate their performance. In~\cref{sec:e_nucleus_results}, we benchmark our model with experimental electron-nucleus data and discuss the Hamiltonian dependence arising from different next-to-next-to-leading order (N$^2$LO) chiral Hamiltonians. We then turn to neutrinos in~\cref{sec:neutrino_nucleus_results}, where both results for the charged-current responses and the flux-averaged total \mbox{$\nu_\mu$-$^{12}$C} cross section are presented. Finally, we conclude and outline further directions in~\cref{sec:conclusion}.

\section{Formalism}
\label{sec:formalism}
For incoming and outgoing leptons with four-momenta $k$ and $k'$, respectively, the inclusive double-differential lepton-nucleus cross section at tree level reads in the laboratory frame \cite{Nieves:2011pp, Benhar:2006wy} 
\begin{equation}
    \frac{d^2\sigma}{dE_{\boldsymbol{k}}'d\Omega_{\boldsymbol{k}}'} = \kappa\frac{|\boldsymbol{k}'|}{|\boldsymbol{k}|}L_{\mu\nu}(k,k')W^{\mu\nu}(q),
    \label{eq:lepton_nucleus_xsec}
\end{equation}
where $q=k-k'=(\omega,\boldsymbol{q})$ is the four-momentum transfer, $k=(E_{\boldsymbol{k}},\boldsymbol{k})$ (and similarly for $k'$), $E_{\boldsymbol{k}}$ is the on-shell energy, and the cosine of the scattering angle is $\cos\theta_{\boldsymbol{k}}' = \hat{\boldsymbol{k}}\cdot\hat{\boldsymbol{k}'}$. 
The lepton tensor $L_{\mu\nu}(k,k')$ is only constrained by lepton kinematics. In the charged-current (CC) case, $L_{\mu\nu}$ contains an additional parity-violating term in contrast to electromagnetic (EM) processes. 
The prefactor $\kappa$ depends on the process,   
\begin{equation}
    \kappa_{\text{EM}} = \frac{\alpha^2}{q^4}, \quad \text{and} \quad \kappa_{\text{CC}} = \frac{G_F^2\cos\theta_C^2}{4\pi^2},
    \label{eq:kappa}
\end{equation}
in the limit where the mediating $W$-boson is integrated out. 

All information about the nuclear dynamics is contained in the hadron tensor $W^{\mu\nu}(q)$, which we model within the impulse approximation (IA) and the LDA. 
As the name indicates, the IA is based on the assumption of a short interaction time between the electroweak probe and the nucleus, compared to the time scale of the nuclear dynamics. It is further assumed that the spatial resolution of the mediating boson, which is of order $1/|\boldsymbol{q}|$, is smaller than typical inter-nuclear distances, but sufficiently large not to resolve the internal structure of the nucleon \cite{Chew:1950xxx, Chew:1952fca}. Consequently, the IA simplifies lepton-nucleus scattering to the incoherent sum of lepton scattering off a single bound nucleon \cite{Benhar:2006wy}. 
The sum over the bound nucleons can be obtained by integrating over the hole SF, $S_h(E,\boldsymbol{p})$, which gives the joint probability of removing a nucleon, with removal energy $E$ and momentum $\boldsymbol{p}$, from the nucleus \cite{Dickhoff:2008:book}. 
In this work, we also account for FSI through the particle SF, $S_p(E,\boldsymbol{p})$, using both the local relativistic Fermi gas (LRFG) and the realistic \emph{ab initio} calculation (see below).

To obtain the hole and particle SFs for different nuclei, we employ the LDA, which is based on the assumption that a finite nucleus can be locally described by infinite NM evaluated at the nuclear density in that point \cite{Sobczyk_phd, Benhar:1994hw, Marino:2021xyd}. 
Under the IA+LDA assumptions, the hadron tensor for a nucleus with density profile $\rho(r)$, where $r=|\boldsymbol{r}|$, is given by \cite{Nieves:2017lij}
\begin{equation}
    \begin{split}
        &W^{\mu\nu}(q) =\int d\boldsymbol{r}\int\frac{d\boldsymbol{p}}{(2\pi)^3}\int dE\, S^{\text{NM}}_{h}(E,|\boldsymbol{p}|,\rho(r))\\
        &\times S^{\text{NM}}_{p}(E+\omega,|\boldsymbol{p}+\boldsymbol{q}|,\rho(r))\frac{1}{4E_{\boldsymbol{p}}E_{\boldsymbol{p}+\boldsymbol{q}}}w^{\mu\nu}(p,q),
        \label{eq:hadron_tensor}
    \end{split}
\end{equation}
where $w^{\mu\nu}(p,q)$ is the single-nucleon tensor and the NM hole and particle SFs, $S_h^{\text{NM}}$ and $S_p^{\text{NM}}$, are rotationally invariant and defined as
\begin{align}
   & S^{\text{NM}}_h(E,|\boldsymbol{p}|,\rho) = S^{\text{NM}}(E,|\boldsymbol{p}|,\rho), \quad E < \mu(\rho), \nonumber\\
   & S_p^{\text{NM}}(E,|\boldsymbol{p}|,\rho) = S^{\text{NM}}(E,|\boldsymbol{p}|,\rho),\quad E>\mu(\rho),
\end{align}
where $\mu(\rho)$ is the chemical potential. In the above equations, we suppressed the isospin dependence to simplify the notation. 

We normalize the proton density $\rho^p(r)$ as $\int d\boldsymbol{r}\rho^p(r) = Z$,
and assume that for isospin-symmetric nuclei ($N=Z$) the neutron density is the same as the proton density, i.e., $\rho^n(r)=\rho^p(r)$. Furthermore, the normalization of the NM SF is \cite{FrancescoMarino:2026dzw}
\begin{equation}
    \int_{-\infty}^{+\infty} dE\,S^{\text{NM}}(E,|\boldsymbol{p}|,\rho) = 1,
    \label{eq:NM_SF_sum_rule}
\end{equation}
for fixed $|\boldsymbol{p}|$ and $\rho$. Note that the symmetric NM density $\rho=2\rho^p$. The NM SF should also fulfill the self-consistent relation
\begin{equation}
    \rho = g\int \frac{d\boldsymbol{p}}{(2\pi)^3} n(|\boldsymbol{p}|,\rho),
    \label{eq:density_mom_distr}
\end{equation}
where $g=4$ is a degeneracy factor (two for spin and two for isospin) and the occupation number is 
\begin{equation}
    n(|\boldsymbol{p}|, \rho) = \int_{-\infty}^{\mu(\rho)} dE\,S^{\text{NM}}(E,|\boldsymbol{p}|, \rho).
    \label{eq:occupation_number}
\end{equation}

Since the \emph{ab initio} SFs from SCGF are non-relativistic, they become unreliable at high momentum and energy transfers. Therefore, at large $|\boldsymbol{q}|$, we will instead employ the plane wave impulse approximation (PWIA) where the only FSI are Pauli blocking. In that case, $S_p^{\text{NM}}$ in~\cref{eq:hadron_tensor} is given by the corresponding particle SF of the LRFG,
\begin{equation}
    \begin{split}
        S^{\text{LRFG}}_p(E,|\boldsymbol{p}|,\rho) &=\delta(E+m_N-E_{\boldsymbol{p}})\Theta(|\boldsymbol{p}|-p_F(\rho)),
        \label{eq:particle_SF_LRFG}
    \end{split}
\end{equation}
where $p_F(\rho)$ is the Fermi momentum and $\Theta$ is the Heaviside step function. We will thus consider two models for the hadron tensor that are expected to work in different kinematical regimes. We refer to them as the PWIA and IA+FSI:
\begin{enumerate}
    \item IA+FSI: The hole and particle SFs are from the \emph{ab initio} NM calculation. 
    \item PWIA: The hole SF is from the \emph{ab initio} NM calculation, while the particle SF is given by~\cref{eq:particle_SF_LRFG}.
\end{enumerate} 

The single-nucleon tensor can be written as
\begin{equation}
    w^{\mu\nu}(p,q) = \sum_{\text{spins}}\bar{u}(p')j^\mu(q)u(p)[\bar{u}(p')j^\nu(q)u(p)]^\dagger,
\end{equation}
where $p$ ($p'$) is the four-momentum of the initial (final) nucleon, the Dirac spinors are normalized as $\bar{u}(p)u(p)=2m_N$, where $m_N$ is the nucleon mass, and 
$j^{\mu}$ is the current for the process under consideration.
We note that for the PWIA approach, we will follow prescription used by Benhar~\cite{Benhar:2006wy} which introduces a modified momentum transfer in the single-nucleon tensor $w^{\mu\nu}(p,q)\rightarrow w^{\mu\nu}(p,\Tilde{q})$, $\Tilde{q}=(\Tilde{\omega},\boldsymbol{q})$, where $\tilde{\omega}= m_N + E + \omega - E_{\boldsymbol{p}}$. In the IA+FSI case, we set $\tilde{\omega} = E_{\boldsymbol{p}+\boldsymbol{q}}-E_{\boldsymbol{p}}$. Additionally, we require $\Tilde{\omega}>0$.

The nucleon-nucleon vertices entering $w^{\mu\nu}$ can be parametrized in terms of form factors, see, e.g., Ref.~\cite{Thomas_Weise:2001} for explicit expressions. Due to the conserved vector current (CVC) for EM processes, only two form factors are needed. For CC processes, we assume the CVC and the partially conserved axial vector current (PCAC) hypotheses, which result in the additional axial and pseudoscalar nucleon form factors. Unless otherwise specified, the parametrization by Galster \emph{et al.} \cite{Galster:1971kv} is employed for the vector form factors, we use the dipole approximation for the axial nucleon form factor with $M_A=1.049$ GeV \cite{Nieves:2004wx}, and assume pion-pole dominance for the pseudoscalar form factor \cite{Thomas_Weise:2001}. 

Nuclear response functions can be defined from the components of the hadronic tensor in \cref{eq:hadron_tensor}.
For EM processes, the responses $R_L$ and $R_T$, corresponding to longitudinally and transversely polarized photons, are given by 
 \cite{Benhar:2006wy}
\begin{equation}
    \begin{split}
        R_L(q) &=W^{00}(q), \quad \text{and}\\
        R_T(q) &=(\delta_{ij} - \frac{q_iq_j}{\boldsymbol{q}^2})W^{ij}(q).\\
    \end{split}
\end{equation}
For CC processes, there are five independent responses given by the ${00}$, ${0z}$, ${zz}$, ${xx}$, and ${xy}$ components of $W^{\mu\nu}$ when setting the momentum transfer along the $z$-direction~\cite{Nieves:2011pp}. 

\section{Spectral functions for nuclear matter}
\label{sec:data}

We determine NM SFs using the Gorkov SCGF approach of Refs.~\cite{FrancescoMarino:2026dzw,Marino:2024tfp}.
The Gorkov framework naturally incorporates pairing correlations, making it suitable for open-shell nuclei~\cite{Soma:2011aj,Barbieri:2021ezv,Soma:2020xhv} and, in the NM case, rendering the method remarkably stable across density regimes, even in the low-density superfluid phase~\cite{Marino_superfluid}.

In homogeneous matter, single-particle properties are diagonal in the momentum $\mathbf{p}$, isospin projection $m_t=p,n$, 
and spin projection $m_s= \uparrow, \downarrow$. Moreover, isotropy implies that they are functions of $\abs{\mathbf{p}}$. Also, since we consider spin-saturated matter, we solve for one spin projection only.
The single-nucleon SF is represented by a discrete number of pairs of peaks, which are located symmetrically with respect to the chemical potential, namely,
\begin{align}
    \label{eq: SF ADC}
    S^{\text{NM}}_{m_tm_s}(\omega,|\boldsymbol{p}|,\rho) &= \sum_{i} \bigg[ \abs{ \mathcal{V}^{i}_{\SpVars} }^{2} \delta(\omega + \omega_i)
    \\ & + \abs{ \mathcal{U}^{i}_{\SpVars} }^{2} \delta(\omega - \omega_i) \bigg]\delta_{|\boldsymbol{p}|,p_i}\delta_{\rho,\rho_i},
    \nonumber
\end{align}
and is fully characterized by the energy poles $\omega_i$, denoting the energy of the excited state $\ket{\Psi_i}$ with respect to the ground-state energy $E_0$. The corresponding amplitudes for the addition and removal of a nucleon read $\mathcal{U}_{\SpVars}^{i} = \mel{\Psi_0}{c_{\SpVars}}{\Psi_i}$ and $\mathcal{V}_{\SpVars}^{i} = \mel{\Psi_0}{c_{\SpVars}^{\dagger}}{\Psi_i}$, respectively.
We stress that a different SF is obtained at each density $\rho$, and drop from now on the subscript $m_s$. 

A finite number of solutions $i$ is obtained by solving the Gorkov equations, which in the implementation of Refs.~\cite{Marino:2024tfp,FrancescoMarino:2026dzw} can be cast as a self-consistent eigenvalue problem. Denoting
\begin{equation}
\label{eq:tilde_t}
    \tilde{t}_{\SpIso}(\mathbf{p},\omega) \equiv
  t_{\mathbf{p}} - \mu_{\SpIso} + \Sigma^{11}_{\SpIso}(\mathbf{p},\omega),
\end{equation}
we have
\begin{equation}
    \label{eq: Gorkov energy dependent}
\begin{pmatrix}
  \tilde{t}_{\SpIso}(\mathbf{p},\omega_i) & \SigAnom_{\SpIso}(\mathbf{p}) \\
  \SigAnom_{\SpIso}(\mathbf{p}) & -\tilde{t}_{\SpIso}(\mathbf{p},-\omega_i)
\end{pmatrix}
\begin{pmatrix} \mathcal{U}^{i}_{\SpMomIso} \\ \mathcal{V}^{i}_{\SpMomIso} \end{pmatrix}
= \omega_i
\begin{pmatrix} \mathcal{U}^{i}_{\SpMomIso} \\ \mathcal{V}^{i}_{\SpMomIso} \end{pmatrix} .
\end{equation} 
In Eqs.~\eqref{eq:tilde_t} and~\eqref{eq: Gorkov energy dependent}, $t_{\mathbf{p}} =  \mathbf{p}^{2}/(2m_N)$ is the kinetic energy, $\mu_{\SpIso}$ is the chemical potential for the appropriate fermion species, $\SigAnom_{\SpIso}(\mathbf{p})$ is the pairing field, which we approximate at leading order (i.e., it has no energy-dependence), and $\Sigma^{11}_{\SpIso} (\mathbf{p},\omega)$ is the normal self-energy. 
Many-body correlations are encoded in the self-energy, which we treat using the algebraic diagrammatic construction (ADC) method~\cite{Barbieri:2016uib,FrancescoMarino:2026dzw,Barbieri:2021ezv}.
In particular, all results in this work are based on the so-called ``ADC(3)-D'' truncation~\cite{Marino:2024tfp,Barbieri:2016uib}, which extends the third-order ADC approximation, ADC(3), by incorporating into the self-energy the two-particle-two-hole amplitudes from a preliminary coupled-cluster computation~\cite{Hagen:2013yba} at the doubles level (hence the ``D'').
The ADC(3)-D truncation includes the infinite classes of the ladder diagrams and ring diagrams, among other contributions.
Thus, it accounts in a fully microscopic way for the medium-correction effects discussed in Ref.~\cite{Nieves:2017lij} and references therein and achieves state-of-the-art accuracy on the NM equation of state (EOS)~\cite{Marino:2024tfp,Hu:2025udi}.

Within the ADC framework, we simulate infinite matter using a finite number of nucleons $A$ enclosed in a box~\cite{LietzCompNucl,FrancescoMarino:2026dzw}.
We enforce homogeneity by imposing special-point twisted-angle boundary conditions (sp-TABC)~\cite{Hagen:2013yba,FrancescoMarino:2026dzw}. 
While single-particle momenta $\mathbf{p}$ are discrete in this framework, sp-TABCs provide a sufficiently dense mesh of momentum points, much finer than that of standard periodic boundary conditions.
We use $A=132$ as it allows to minimize finite-size effects on the EOS.
The computational cost for each density point is about 3000 CPUh.

We introduce a convenient notation for the SF from Eq.~\eqref{eq: SF ADC} as a function of the energy $E = \mu_{\SpIso} + \omega$ by writing down
\begin{align}
    \label{eq:NM_SF_delta_weights}
    S^{\text{NM}}_{(\mathscr{H},\SpIso)}(E,\abs{\mathbf{p}},\rho) = \sum_{i} \mathcal{A}_{i}\delta(E-\epsilon_i)\delta_{|\boldsymbol{p}|,p_i}\delta_{\rho,\rho_i} \delta_{m_t, m_{t_i} },
\end{align}
where the energies $\epsilon_i$ and dimensionless delta weights $\mathcal{A}_{i}$ are defined by associating to each solution $\omega_i$, $\mathcal{V}^{i}_{\SpVars}$, $\mathcal{U}^{i}_{\SpVars}$ of the Gorkov equations, respectively, the particle contributions $\epsilon_i = \mu_{\SpIso} + \omega_{i}$ and $\mathcal{A}_i = \abs{\mathcal{U}^{i}_{\SpVars}}^{2}$, and the hole contributions $\epsilon_i = \mu_{\SpIso} - \omega_{i}$ and $\mathcal{A}_i = \abs{\mathcal{V}^{i}_{\SpVars}}^{2}$.
The weight $\mathcal{A}_{i}$ satisfy the normalization $\sum_i \mathcal{A}_{i} = 1$ for any momentum state.
Also, we have specified the dependence on the nuclear Hamiltonian by the subscript $\mathscr{H}$.

\begin{figure}[h]
    \centering
    \includegraphics[width=\columnwidth]{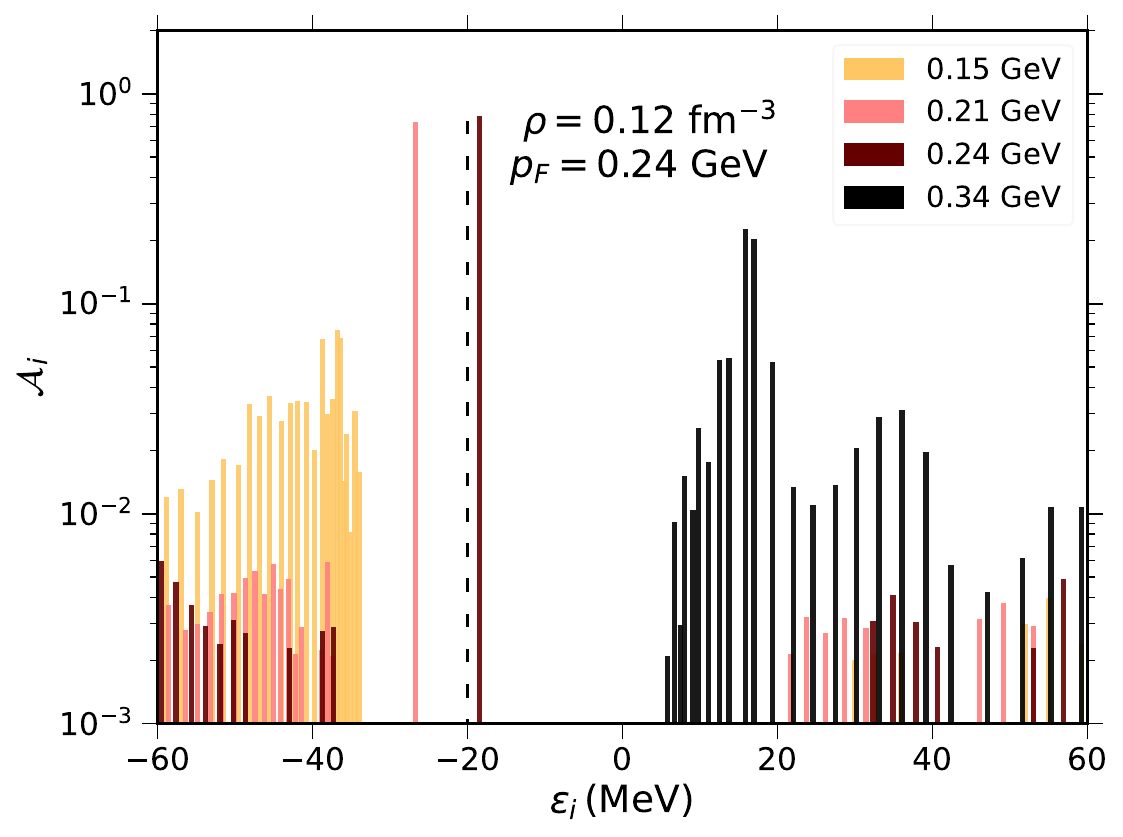}
    \caption{SF obtained with the $\Delta \rm{NNLO_{GO}}(450)$ interaction~\cite{Jiang:2020the} at density $\rho = 0.12$ fm$^{-3}$, corresponding to a Fermi momentum $p_F=0.24$ GeV.
    The weights $\mathcal{A}_i$ of the SF are shown as a function of the corresponding energies $\epsilon_i$ for four different momenta (see text). The dashed vertical line is the chemical potential.
    }
    \label{fig: SpectralFunction_Slices}
\end{figure}

In Fig.~\ref{fig: SpectralFunction_Slices}, we show an example of NM SF derived with SCGF (see also~\cite{FrancescoMarino:2026dzw}).
The spectral weights $\mathcal{A}_i$ and the corresponding energies $\epsilon_i$ are shown for four different momenta spanning $0.15-0.34$ GeV.
There is a distinctive qualitative difference between the spectral distribution of states with momenta close to the Fermi momentum $p_F$ (marked in light and dark red in Fig.~\ref{fig: SpectralFunction_Slices}) and that of states further apart from $p_F$ (plotted in yellow and black).
In the first case, the SF appears concentrated in a single peak close to the Fermi energy (dashed vertical line) exhausting a significant fraction of the strength (roughly 70-80\%).
In the other case, no prominent peak emerges. Rather, the SF is distributed over many minor structures. This strength redistribution is a consequence of the inclusion of dynamical correlations in Eq.~\eqref{eq: Gorkov energy dependent} through an energy-dependent self-energy.
Note that, in the free Fermi gas or in the Hartree-Fock mean-field approximation, the SF is characterized by a unique peak at each momentum~\cite{Barbieri:2016uib}, thus missing fragmentation effects.

\section{Neural network setup}
\label{sec:ANN}
The SCGF calculation described in \cref{sec:data} produces SFs at discrete points in energy, momentum and density while the calculation of cross sections requires the SFs at arbitrary values of these variables.
We exploit the capabilities of neural networks to determine a multi-dimensional interpolation of the \textit{ab initio} SFs.
Related studies include, e.g., Refs.~\cite{Dornheim2019,Dong2024,martinezmarimon2024}.
In particular, we develop six models, one for each combination of Hamiltonian $\mathscr{H}\in \{\Delta\text{N}^2\text{LO}_{\text{GO}}(394)$, $\Delta\text{N}^2\text{LO}_{\text{GO}}(450)$, $\text{N}^2\text{LO}_{\text{sat}}(450)\}$ and isospin $\SpIso\in\{p,n\}$.
The $\Delta$-full GO models are detailed in Ref.~\cite{Jiang:2020the}, while $\text{N}^2\text{LO}_{\text{sat}}(450)$ is from Ref.~\cite{Ekstrom:2015rta}.

\subsection{Data pre-processing}
\label{subsec:ANN:data_preprocessing}
The NM SFs from the SCGF method are represented by Eq.~\eqref{eq:NM_SF_delta_weights}.
From now on, we will suppress the $(\mathscr{H},\SpIso)$ subscript for the sake of simplicity. 
The SCGF data contain NM densities in the interval $\rho\in[0.00125, 0.2]$ fm$^{-3}$. When integrating over the nucleus volume in Eq.~\eqref{eq:hadron_tensor}, we set the upper boundary of the density-normalization integral to $r_{\text{max}} = \text{argmin}(\rho(r) - 0.00125\text{ fm$^{-3}$})$. This results in a negligible deviation of around 0.8\% from the exact number of protons. Furthermore, the momentum range of the data depends on the density. For the lowest density, $|\boldsymbol{p}|\in [0.01, 0.14]$ GeV, while for the highest density, $|\boldsymbol{p}|\in[0.08, 0.76]$ GeV. 

The ADC-SCGF scheme relies on a discrete representation of the SF as a superposition of Dirac-delta peaks.
We convert the delta weights into a smooth distribution along the energy dimension by applying a Lorentzian folding
\begin{equation}
    \begin{split} 
        S^{\text{NM}}(E, |\boldsymbol{p}|, \rho, \Gamma)&=\frac{\Gamma}{\pi}\sum_{i}\frac{\mathcal{A}_i}{(E-\epsilon_i)^2+\Gamma^2}\delta_{|\boldsymbol{p}|,p_i}\delta_{\rho,\rho_i},\\
        \label{eq:NM_SF_Lorentzian_smoothing}
    \end{split}
\end{equation}
where $\Gamma$ is a smoothing factor in units of energy. 
This is physically well-motivated and is also technically convenient, as it allows to augment the dimension of the SF dataset by evaluating Eq.~\eqref{eq:NM_SF_Lorentzian_smoothing} on an energy mesh.
Note that $S^{\text{NM}}$ in \cref{eq:NM_SF_Lorentzian_smoothing} now also becomes a function of the auxiliary variable $\Gamma$. 
For each $(|\boldsymbol{p}|, \rho)$ we sample $N_E=1500$ points in energy. We take 50\% of the points linearly distributed in the interval $E\in[-0.4,0.4]$ GeV and 50\% of the points sampled with inverse transform Monte Carlo sampling; see, e.g., Chapter 42 of Ref. \cite{PDG_Navas:2024}. This results in more data points around the dominating SF peaks while still covering the entire $E$ region. 

The energy-smoothed data, for a specific Hamiltonian and isospin, are stored as
\begin{equation}
    \mathcal{D}_{(\mathscr{H},\SpIso)} = \{\boldsymbol{X}, \boldsymbol{S}^{\text{NM}}_{\text{true}}\},
\end{equation}
where the feature matrix $\boldsymbol{X}$ is $N_{\mathcal{D}}\times 4$ dimensional with rows $\boldsymbol{X}_{i,:} = [E_i,p_i, \rho_i, \Gamma_i]$, $i=1,...,N_{\mathcal{D}}$, and the $i$:th element of the true output vector corresponds to $S^{\text{NM}}_{\text{true}}(\boldsymbol{X}_{i,:})$. For each Hamiltonian and isospin, $N_\mathcal{D}\approx 10^7$. The feature columns $\boldsymbol{X}_{:,j}$, $j=1,...,4$, are rescaled according to the standard score
\begin{equation}
    \boldsymbol{X}_{:,j} = \frac{\boldsymbol{X}_{:,j} - \bar{\boldsymbol{X}}_{:,j}}{\sigma_j},
    \label{eq:standard_score}
\end{equation}
where $\bar{\boldsymbol{X}}_{:,j}$ and $\sigma_j$ are the mean and standard deviation of column $j$, respectively. 
The rescaling in \cref{eq:standard_score} results in an equal treatment of the features during training. 

\subsection{Neural network architecture, loss function and hyperparameters}
\begin{figure}
    \centering
    \includegraphics[width=\linewidth]{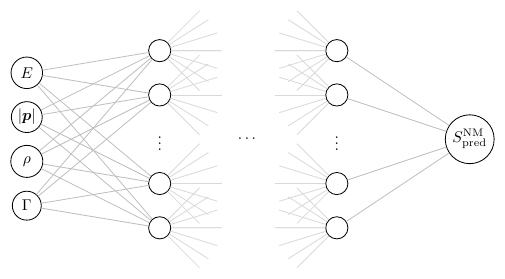}
    \caption{Schematic sketch of the neural network architecture showing the four input nodes $E$, $|\boldsymbol{p}|$, $\rho$ and $\Gamma$, several fully connected hidden layers, and the output layer $S^{\text{NM}}_{\text{pred}}(E,|\boldsymbol{p}|, \rho, \Gamma)$.}
    \label{fig:ANN_architecture}
\end{figure}

We construct a fully connected, convolutional neural network for each Hamiltonian and isospin using TensorFlow and Keras \cite{tensorflow2015-whitepaper}.  The input layer consists of four nodes corresponding to a row of the input feature matrix $\boldsymbol{X}_{i,:}$, and the output layer contains a single node corresponding to the predicted NM SF $S^{\text{NM}}_{\text{pred}}(\boldsymbol{X}_{i,:})$. Between the input and output layers, we use four hidden layers with 16, 32, 64, and 128 nodes.  For a schematic sketch, see \cref{fig:ANN_architecture}. 
In order to tune some of the hyperparameters and evaluate the performance of our neural networks, we split each dataset $\mathcal{D}_{(\mathscr{H},\SpIso)}$ into 70\% train, 20\% validation and 10\% test data. At this stage, we keep all $N_E$ data points belonging to a fixed value of $|\boldsymbol{p}|$, $\rho$, and $\Gamma$ within the same set (train, validation or test). This choice is motivated below. In the following, the set of data points in which only $E$ and $S^{\text{NM}}$ vary will be referred to as a \emph{sample}.  

The neural networks are trained in two phases. In the first 70 epochs, we use the mean absolute error (MAE) loss function, which is robust and non-sensitive to outliers. Then, we train for an additional 30 epochs using a modified loss function with a physics-informed penalty term. The penalty term encodes the fulfillment of the sum rule in \cref{eq:NM_SF_sum_rule}.
The modified MAE, denoted MAE$_{\text{mod}}$, reads
\begin{equation}
    \text{MAE}_{\text{mod}} = \text{MAE} + \eta |\int_{-\infty}^\infty dE\, S^{\text{NM}}_{\text{pred}} - 1|,
    \label{eq:MAE_mod}
\end{equation}
where the weight factor $\eta$ is set so that the penalty term is approximately 10\% of the MAE. The integration over $E$ in the second training phase requires that each \emph{sample} remain within a single batch. We do not wish to change the hyperparameters between the different phases. Therefore, we set the batch size to $N_E$. To ensure robust training, the training data are fully shuffled in the first phase, and the \emph{samples} are shuffled in the second phase. 

In the input and hidden layers we use the leaky relu activation function \cite{leaky_relu}, and in the output layer we use softplus activation \cite{softplus} to ensure positivity. We employ the adaptive moment estimation (Adam) optimizer \cite{adam_optimizer} and tune the optimizer hyperparameters (the learning rate $\eta$ and the moments $\beta_1$ and $\beta_2$) using Keras built-in Bayesian optimization package. Different combinations of Hamiltonian and species result in similar but slightly different values of $\eta$, $\beta_1$, and $\beta_2$. For consistency we choose the median values $\eta=0.0001$, $\beta_1=0.85$, and $\beta_2=0.999$ in all neural networks. 

\subsection{Neural network performance}
\begin{figure}[htbp]
    \centering
    \includegraphics[width=\linewidth]{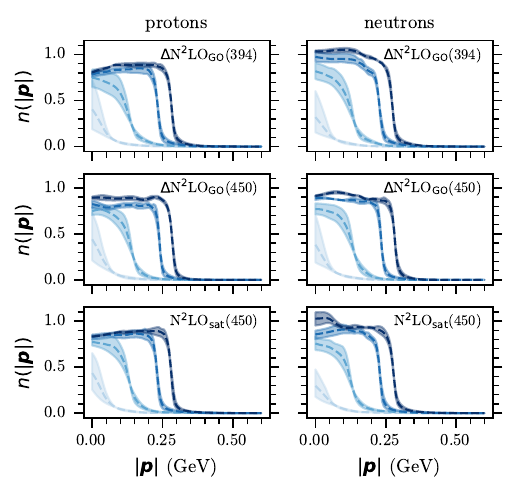}
    \caption{NM occupation number $n(|\boldsymbol{p}|)$ predicted by the neural network, see \cref{eq:occupation_number}, as a function of the momentum $|\boldsymbol{p}|$ for densities $\rho\in \{0.00125, 0.02, 0.1, 0.18\}$ fm$^{-3}$. Lighter colors correspond to lower densities. 
    Left (right) panels refer to proton (neutron) occupations, and each row corresponds to a different interaction.
    The uncertainty bands come from the variation of $\Gamma\in\{1.0,2.0,3.0\}$ MeV.}
    \label{fig:momentum_distribution_ANN}
\end{figure}

Having obtained our trained neural networks, we calculate the chemical potential, which enters the definition of the hole and particle SFs, by checking the self-consistency of the relation in Eq.~\eqref{eq:density_mom_distr}. 
This is done separately for each Hamiltonian, isospin, and for varying values of $\Gamma$. In \cref{fig:momentum_distribution_ANN} we show the obtained distributions $n(|\boldsymbol{p}|,\rho)$, for the lowest and highest NM densities probed in our calculations, and for two values in between. The uncertainty bands in the figure come from varying $\Gamma\in\{1.0,2.0,3.0\}$ MeV. 

The occupation number distributions in~\cref{fig:momentum_distribution_ANN} are quite similar for the various Hamiltonians. The main differences are in the low-momentum region for the highest density. It should be highlighted that the training data do not contain momenta down to $|\boldsymbol{p}|=0$ and therefore, the networks are less reliable in that region. However, since $n(|\boldsymbol{p}|)$ is multiplied by $\boldsymbol{p}^2$ in the cross-section calculations (see Eq.~\ref{eq:hadron_tensor}), these differences will not influence the physical observables. 

\begin{figure*}[b]
    \centering
    \includegraphics[width=\linewidth]{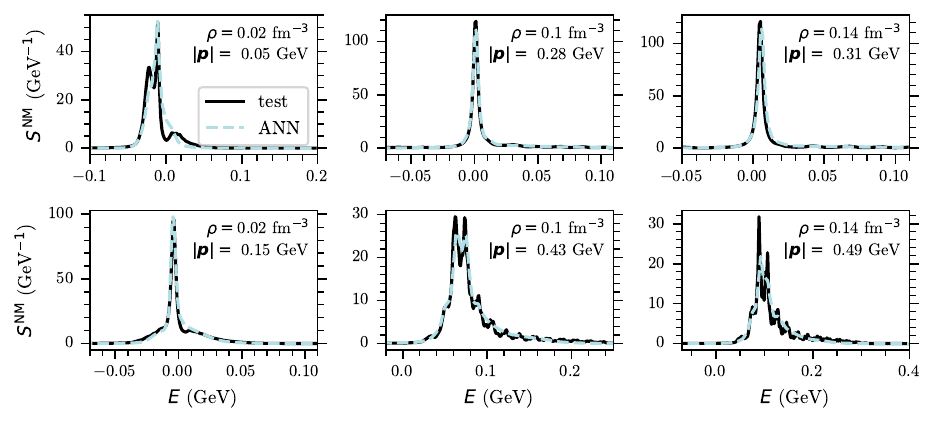}
    \caption{Representative set of neural network predictions (dashed line) vs. test data (solid line) for neutrons using the N$^2$LO$_{\text{sat}}$(450) potential and $\Gamma=2.0$ MeV.}
    \label{fig:ANN_predictions}
\end{figure*}

We next evaluate the performance of our neural networks by looking at a representative set of predicted SFs, compared to data from the test set. This can be seen in \cref{fig:ANN_predictions}. In general, the networks capture the broad details of the SF from low to high densities and across the range of momenta. As expected, more fragmented SFs are harder to predict. Lastly, we note that applying the modified MAE with the penalty term in the last training phase, allows us to keep the normalization condition of~\cref{eq:NM_SF_sum_rule} up to an error of at most $2\%$. 

\section{Electron scattering results}
\label{sec:e_nucleus_results}
We now proceed to present results for electron-nucleus scattering. In all figures, we include two bands of uncertainty. The outer band (light blue/green) is the full uncertainty obtained by varying the nuclear Hamiltonian and the smoothing factor $\Gamma\in\{1.0,2.0,3.0\}$ MeV, and the dashed line is the mean of this set. The inner band (darker blue/green) corresponds to the uncertainty from only varying the Hamiltonian and keeping $\Gamma=2.0$ MeV fixed. We include results for both the PWIA and the IA+FSI models. For $^{12}$C and $^{16}$O, we use a modified harmonic oscillator parametrization of the charge density and a two-parameter Fermi distribution for $^{40}$Ca. The charge-distribution parameters are from Ref. \cite{DEJAGER1974479}, which provides accurate parametrizations of experimental measurements, and the finite size of the nucleon is taken into account according to Ref. \cite{Garcia-Recio:1991ocp}.   

\begin{figure*}[t]
    \centering
    \includegraphics[width=\linewidth]{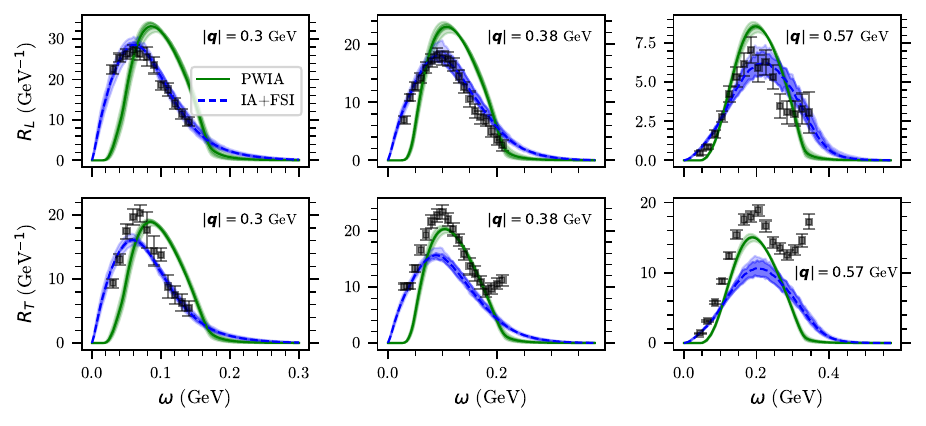}
    \caption{$^{12}$C EM response functions using the PWIA and the IA+FSI models. See main text for details regarding the uncertainty bands. The experimental data are the Rosenbluth separated data from Ref.~\cite{Jourdan:1996np}.}
    \label{fig:12C_response}
\end{figure*}

In~\cref{fig:12C_response}, the electromagnetic response functions of $^{12}$C for three different momentum transfers are shown and compared to the Rosenbluth-separated data from Ref.~\cite{Jourdan:1996np}. 
When we compare our full model, IA+FSI, to the experimental data in the longitudinal response, $R_L$ (top row), we see a remarkable agreement at $\omega \lesssim  \omega_{\text{QE}}$ for all momentum transfers, while we slightly overestimate the strength above the QE peak for the two higher momentum transfers. This is partially due to the non-relativistic nature of the particle SF as discussed earlier. We expect relativistic corrections to the particle SF to redistribute strength from high towards lower $\omega$~\cite{Nieves:2004wx,Nieves:2017lij}.
Neglecting FSI (results denoted as PWIA in~\cref{fig:12C_response}) leads to some discrepancies with the data clearly seen in the longitudinal response for the two lower momentum transfers; the position of the QE peak is shifted towards higher energies by $15-25$ MeV and overestimated by $20-25\%$. At the highest momentum transfer shown, $|\boldsymbol{q}|=0.57$ GeV, the position of the QE peak is better captured by the PWIA model, which indicates that FSI play a less prominent role with the growing momentum.

In the case of the transverse response, $R_T$, theoretical calculations are instead underestimating the experimental data systematically. 
However, this is expected since we only include one-body currents, while the transverse response receives a significant contribution from two-body correlations, see, e.g., \cite{Donnelly:1978xa, Lovato:2015qka}. Otherwise, the behaviour of both IA+FSI and PWIA results is analogous to the $R_L$ case.

The sizes of the theoretical uncertainties included in our calculations are small for both models at the two lower momentum transfers, not exceeding $10\%$ at the QE peak, while the uncertainty grows to around $15\%$ for the IA+FSI model at $|\boldsymbol{q}|=0.57$ GeV. 
It should be highlighted that this larger uncertainty should not be interpreted as stemming solely from the Hamiltonian dependence. Rather, it is also impacted by the uncertainty due to the neural network extrapolations, as no training data at high momenta are available.

\begin{figure}[h]
    \centering
    \includegraphics[width=\columnwidth]{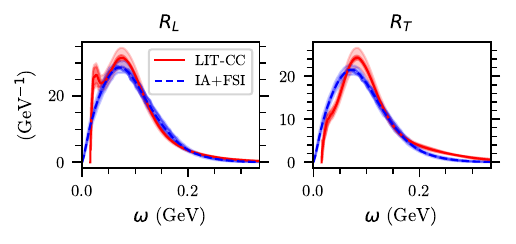}
    \caption{$^{16}$O EM response functions using the IA+FSI model, compared to the LIT-CC calculation of Ref. \cite{Acharya:2024xah}. The momentum transfer is $|\boldsymbol{q}|=0.335$ GeV.}
    \label{fig:16O_response}
\end{figure}

Next, we consider nuclear responses of $^{16}$O. Since the experimental data are very scarce, we compare our results to \emph{ab initio} responses obtained with coupled-cluster theory using Lorentz integral transform (LIT-CC)~\cite{Bacca:2013dma,ChapterBacca2026} of Ref.~\cite{Acharya:2024xah}, see \cref{fig:16O_response}.
\begin{figure*}[t]
    \centering
    \includegraphics[width=\linewidth]{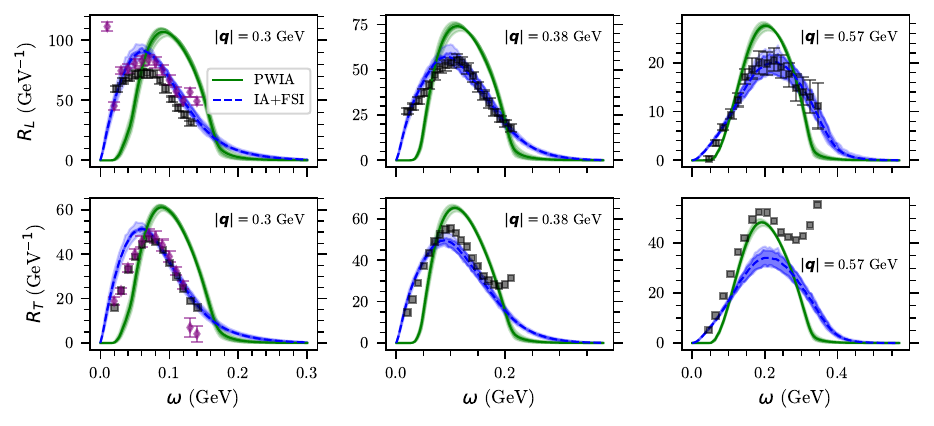}
    \caption{$^{40}$Ca EM response functions using the PWIA and the IA+FSI models. The experimental Rosenbluth separated data are from Ref. \cite{Jourdan:1996np} (squares) and Ref. \cite{Williamson:1997zz} (diamonds).}
    \label{fig:40Ca_set1}
\end{figure*}
The LIT-CC responses employ the $\Delta$N$^2$LO$_{\text{GO}}$(450) Hamiltonian and include uncertainty bands accounting for the $\chi$EFT truncation error. Furthermore, it is a fully non-relativistic calculation, and only one-body currents are included. 
To make the comparison meaningful, we use the same conventions in our calculation, including non-relativistic currents and form factor parameterization.  
The same nuclear dynamics used in both calculations allows us to appreciate the systematic uncertainty coming from our IA+FSI framework. 
Overall, we see a very good agreement between the two calculations in the region of the QE peak, with our prediction slightly quenched and shifted towards low energy transfers. The difference becomes more visible at very low energies. This is to be expected since the LIT-CC calculation naturally offers a more detailed description of the nuclear structure effects which affect mostly the low energy transfer region. 
For example, a rapidly rising ``shoulder'' is observed in the coupled-cluster predictions of the longitudinal response near the proton separation energy ($S_{\text{proton}}(^{16}\text{O}) = 12.12$ MeV), which is imposed when reconstructing the energy-dependent response function from the LIT (see Ref.~\cite{Bacca:2014tla,Acharya:2024xah}). No threshold effect appears in the IA+FSI calculations.

Moving towards medium-mass nuclei, in~\cref{fig:40Ca_set1}, we show $^{40}$Ca responses for the same momentum transfers as in~\cref{fig:12C_response}, compared to the Rosenbluth separated data of Ref.~\cite{Jourdan:1996np} (squares). At $|\boldsymbol{q}|=0.3$ GeV, we also include data from Williamson \emph{et al.} (diamonds) ~\cite{Williamson:1997zz}.  
We note that in the longitudinal response, we find a good agreement between our IA+FSI model and the experimental data, similar in quality to $^{12}$C. The only exception is $|\boldsymbol{q}|=0.3$ GeV where we overestimate the Jourdan data, while being closer to Williamson \emph{et al.}.
In contrast to the $^{12}$C results of \cref{fig:12C_response}, we do not underestimate the transverse response experimental data below $|\boldsymbol{q}|=0.4$ GeV. Rather, our IA+FSI model slightly overestimates or reproduces the correct strength at $|\boldsymbol{q}|=0.3$ GeV, which is somewhat surprising, since we are expecting missing strength due to the lack of two-body currents. Our findings are therefore confirming previous results~\cite{Sobczyk:2023sxh,Franco-Munoz:2022jcl}. 
A qualitatively similar picture persists for other sets of Rosenbluth separated data from Ref.~\cite{Williamson:1997zz} in the range $0.3-0.4$ GeV.

\begin{figure}[t]
    \centering
    \includegraphics[width=\linewidth]{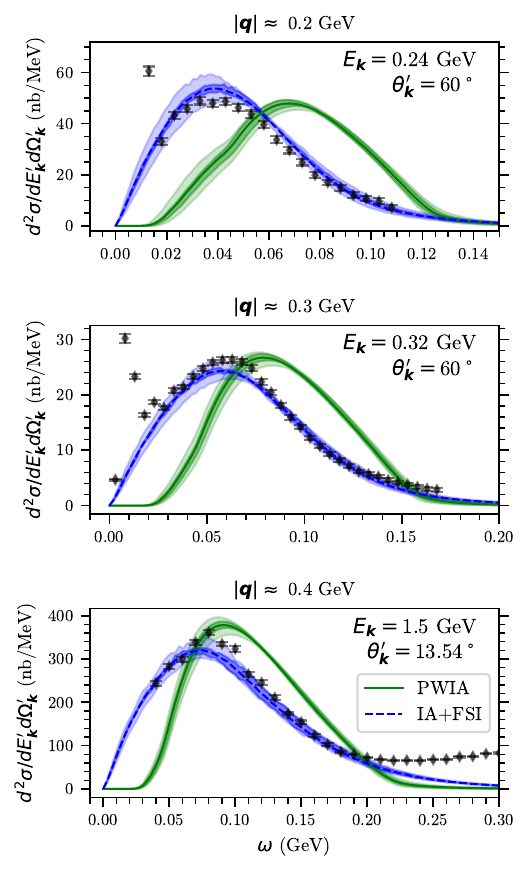}
    \caption{$e^--^{12}$C double-differential cross sections for different kinematical setups, using the PWIA and the IA+FSI models. The approximate momentum transfer probed at each kinematics is indicated in the title of each panel. The experimental data are from Ref.  \cite{Benhar:2006er}}
    \label{fig:12C_xsec1}
\end{figure}

\begin{figure}[t]
    \centering
    \includegraphics[width=\linewidth]{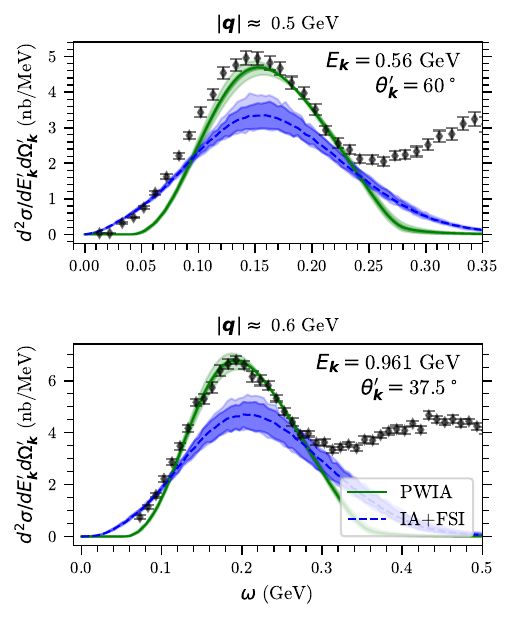}
    \caption{Same as~\cref{fig:12C_xsec1}, but for higher momentum transfers.}
    \label{fig:12C-xsec2}
\end{figure}

To conclude our benchmark against electron data, we present results for the $^{12}$C double differential cross section in~\cref{fig:12C_xsec1} and~\cref{fig:12C-xsec2} for five kinematics, corresponding to different momentum transfers. By directly comparing with cross-sectional data, we circumvent possible ambiguities and systematic uncertainties arising in the Rosenbluth separation. 
At momentum transfer $|\boldsymbol{q}|\approx0.2$ GeV, see the upper panel of~\cref{fig:12C_xsec1}, the IA+FSI model agrees very well with the experimental data, while the PWIA is displacing the peak by around $40$ MeV. When the momentum transfer is increased, $|\boldsymbol{q}|\sim 0.3-0.4$ GeV, as in the middle and lower panel of \cref{fig:12C_xsec1}, we also obtain very good agreement with experiment and a gradual improvement of the PWIA model. 
As expected, the low-lying strength visible in the data is not described by our model. Furthermore, we underestimate the high-energy tail of the data which is due to the fact that mechanisms beyond QE scattering, such as pion production, start contributing.

When the momentum transfer is further increased, see \cref{fig:12C-xsec2}, we note that for $|\boldsymbol{q}|\approx0.5$ GeV, our IA+FSI model describes the data well below the QE peak, where our uncertainty is still small, while the PWIA reproduces the data almost perfectly at and above the peak up to the dip region. At this momentum transfer, the large uncertainty of the IA+FSI model points that we are at the verge of its applicability. On the other hand, the good agreement between the data and the PWIA in both panels of \cref{fig:12C-xsec2}, might be misleading. At these $|\boldsymbol{q}|$, the contribution from the transverse response is large \cite{Sobczyk:2017mts}, and we expect an enhancement at the QE peak when accounting for additional mechanisms.  

\section{Neutrino scattering results}
\label{sec:neutrino_nucleus_results}
\begin{figure}[t]
    \centering
    \includegraphics[width=\linewidth]{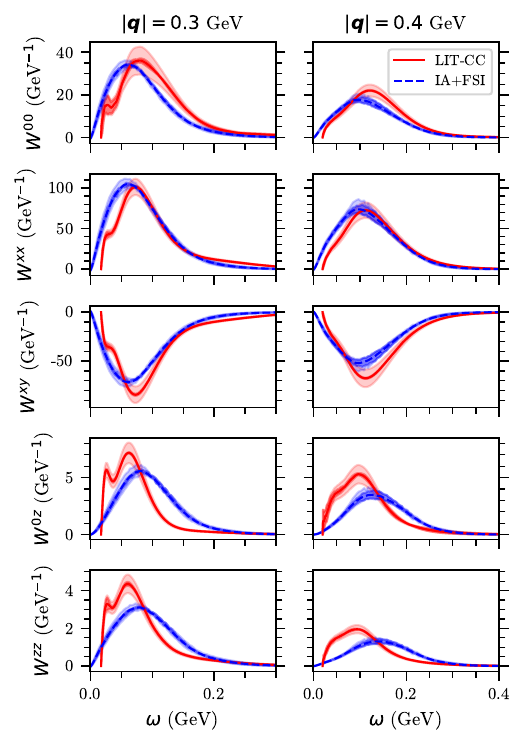}
    \caption{$^{16}$O CC response functions using the IA+FSI model, compared to the LIT-CC calculation of Ref. \cite{Acharya:2024xah}.}
    \label{fig:CC_16O_response}
\end{figure}

Due to the limited cross-section data available for charged-current processes, in Fig.~\ref{fig:CC_16O_response} we first compare the predictions of our model with available LIT-CC theoretical calculations. 
Similarly to the electron case shown in~\cref{fig:16O_response}, the comparison is meaningful since both approaches use the same description of nuclear dynamics and we expect the main differences between them to stem from our approximations (IA and LDA).
First of all, let us focus on $W^{00}, W^{xx}$, and $W^{xy}$ responses whose strength exceeds that of $W^{zz}$ and $W^{0z}$ and which therefore will dominate the cross-section predictions. In this case, our results are qualitatively similar to the electron-scattering comparison shown in~\cref{fig:16O_response}. We predict a similar shape and strength of the QE peak as LIT-CC, with a shift towards lower energies, while we do not recover the threshold behavior of the LIT-CC results. For $W^{zz}$ and $W^{0z}$, the peak is instead displaced towards higher energies and our calculation predicts a more symmetric shape than the LIT-CC calculation. The different behavior in the $W^{0z}$ and $W^{zz}$ responses is probably due to the fact that the pseudoscalar form factor, $F_P$, enters these responses\footnote{The $W^{00}$ response also depends on $F_P$. However, the dependence of $W^{00}$ on $F_P$ is much weaker.}. This form factor becomes large at small $Q^2=-q^2$ and is therefore very sensitive to the exact prescription of the off-shell kinematics.

Lastly, we would like to compare our model to the measurements of the $\nu_\mu-^{12}$C total cross section obtained using the Liquid Scintillator Neutrino Detector (LSND) at the Los Alamos National Laboratory \cite{LSND:1994urh,LSND:1997lta, LSND:2002oco}. Since the incoming neutrino energy in this experiment is not monoenergetic, we calculate the flux-averaged total cross section given by 
\begin{equation}
    \bar{\sigma} = \frac{1}{\mathcal{N}}{\int_{E_{\text{min}}}^{E_{\text{max}}}dE_k \sigma(E_k)\Phi(E_k)},
    \label{eq:flux_averaged_xsec}
\end{equation}
where the normalization factor
\begin{equation}
    \mathcal{N} = \int_{E_{\text{min}}}^{E_{\text{max}}}dE_k\Phi(E_k),
    \label{eq:normalization_flux}
\end{equation}
$\Phi(E_k)$ is the neutrino flux in arbitrary units, and $E_{\text{min}}$ and $E_{\text{max}}$ are the threshold energy and the maximum energy of the incoming neutrino, respectively.  

We take the flux from Ref. \cite{LSND:1994urh}, and the threshold and maximum energies are given by $E_{\text{min}}=0.1237$ GeV and $E_{\text{max}}=0.28$ GeV \cite{LSND:1994urh}. For this flux which is peaked around $E_k\approx75$ MeV, most of the strength lies in the range $|\boldsymbol{q}| \sim 0.1-0.4$ GeV, which corresponds to the region where the IA+FSI model is reliable. Furthermore, we remove non-physical strength at low energy transfers by requiring that $\omega > S_{\text{proton}}(^{12}\text{C})$, where the proton separation energy is given by $S_{\text{proton}}(^{12}\text{C}) = 15.96$ MeV \cite{Mohr:2024kco}. 

The resulting flux-averaged total cross sections, using the three different Hamiltonians, and the two different models (PWIA and the IA+FSI models), are summarized in~\cref{tab:neutrino_total_xsec}. 
\begingroup
\renewcommand{\arraystretch}{1.7}
\setlength{\tabcolsep}{8pt}
\begin{table}[htb]
    \centering
    \caption{Flux-averaged total $\nu_\mu$-$^{12}$C total cross section $\bar{\sigma}_\mu$, see \cref{eq:flux_averaged_xsec}, in units of $10^{-40}$ cm$^2$. We consider both the PWIA and the IA+FSI model, for the three N$^2$LO Hamiltonians. The uncertainty in the theoretical calculations comes from the variation of $\Gamma\in[1.0,2.0,3.0]$ MeV. The uncertainties in the experimental data are statistical and systematic, respectively.}
    \begin{tabular}{c c c} \hline \hline
         &PWIA  & IA+FSI\\ \hline
        $\Delta$N$^2$LO$_{\text{GO}}(394)$ &$7.99\pm0.33$& $10.43\pm0.08$\\ \hline
       $\Delta$N$^2$LO$_{\text{GO}}(450)$ &$8.31\pm0.28$& $10.35\pm0.12$\\ \hline
        N$^2$LO$_{\text{sat}}(450)$  &$7.97\pm0.29$&  $10.40\pm0.16$ \\ \hline \hline
        & Experiment & \\ \hline
        LSND  \cite{LSND:1994urh} & LSND \cite{LSND:1997lta} & LSND \cite{LSND:2002oco} \\
        \hline
         $8.3\pm 0.7\pm1.6$ &$11.2\pm0.3\pm1.8$  &  $10.6\pm0.3\pm1.8$  \\ \hline \hline
    \end{tabular}
    \label{tab:neutrino_total_xsec}
\end{table}
\endgroup
We note that both the PWIA and the IA+FSI results are consistent with the experimental data~\cite{LSND:1994urh, LSND:1997lta, LSND:2002oco}, with PWIA lying $\sim20\%$ below IA+FSI, and that the Hamiltonian dependence is small for both models.
These results can be understood as follows. FSI shifts the position of the QE peak and redistributes strength towards lower energy transfers. In the absence of FSI, the peak is displaced by as much as $40$ MeV at $|\boldsymbol{q}|=0.2$ GeV (see \cref{fig:12C_xsec1}). At even lower momenta, the shift is large enough that the peak is no longer fully contained in the region $\omega<|\boldsymbol{q}|$, and the integrated strength therefore falls below the IA+FSI result.
This also gives rise to larger uncertainties from the variation of $\Gamma$ in the PWIA model. For IA+FSI, integrating over $\omega$ across the entire QE peak makes such differences largely cancel, whereas this is no longer the case when only part of the peak contributes. 

\section{Conclusion and Outlook}
\label{sec:conclusion}
In this work, we have set up a framework for modeling lepton-nucleus scattering in which both the initial nuclear dynamics and the FSI are consistently described by \emph{ab initio} nuclear matter SFs derived with SCGF theory. To this end, we developed artificial neural networks acting as interpolators of the discretized data from the \emph{ab initio} calculation. We benchmarked the model for inclusive electroweak observables on isospin-symmetric target nuclei, using three chiral N$^2$LO Hamiltonians.

We assessed the uncertainty of our calculation stemming both from the nuclear Hamiltonian and from the smoothing procedure applied to the discretized \emph{ab initio} results. At low and intermediate momentum transfers, the error budget is dominated by the smoothing procedure, whereas at higher momenta the leading uncertainty originates from the extrapolation of the neural networks beyond the region covered by the training data, which manifests itself as an amplified Hamiltonian dependence. The interpolation error of the networks is not quantified separately, but it is partly absorbed into the smoothing uncertainty: varying $\Gamma$ smears out precisely those sharp features of the SF that are the hardest for the networks to reproduce, so that the spread obtained in this way also probes the sensitivity to the interpolation. We assume the residual interpolation error to be small compared with the other sources. A dedicated quantification, for instance by replacing the present networks with Bayesian neural networks, would be valuable but lies beyond the scope of this work.
The total uncertainty is momentum-dependent and amounts to around $7-15\%$ at the QE peak for $|\boldsymbol{q}|=0.2$--$0.6$~GeV. Once FSI are included, we found remarkably good agreement with the available experimental data for EM processes and with LIT-CC calculations for momentum transfers $|\boldsymbol{q}|\lesssim0.5$~GeV. At higher $|\boldsymbol{q}|$, where the IA+FSI model breaks down, the PWIA approach becomes better founded theoretically.

A precise assessment of the model's performance across different target nuclei could not be established, owing to the lack of experimental data for $^{16}$O and to ambiguities in the $^{40}$Ca datasets. To compensate for the scarcity of data, we compared with the available LIT-CC results for $^{16}$O, for both EM and CC responses, and found overall good agreement within the range of validity of our approach.

An important extension of the model would be to go beyond isospin-symmetric targets towards $^{40}$Ar, relevant for DUNE.
This would require extending the SCGF calculation to isospin-asymmetric matter. Calculations of SFs have been performed in the past~\cite{Rios:2020oad,Rios:2013zqa} with a different Green's function formalism. We are currently extending the ADC framework in this direction~\cite{AsymAdc}, paving the way for reliable computations of SFs and EW cross-sections based on the present approach. Notice that the neural network interpolator will have to be only slightly modified to account for an additional isospin asymmetry variable.

Let us finally point out that the LDA does not incorporate the shell structure of the target nucleus and is therefore not well suited to describing exclusive $(e,e'p)$ data. This deficiency should, however, be far less relevant for semi-exclusive neutrino data, where the flux folding smears out the details of the initial nuclear state. Such a study is part of our future plans, and our approach is well suited to it, since it allows the two descriptions, IA+FSI and PWIA, to be combined into a single model covering the whole relevant phase space. We note that the framework can be readily interfaced with an intranuclear cascade~\cite{Gil:1997jg}, since both rest on the LDA: the outgoing nucleon, described by the particle SF, is associated with a nuclear density that in turn fixes the position inside the nucleus at which the cascade starts. Moreover, the framework can be extended to include in-medium pion production \cite{Oset:1987re, Hernandez:2013jka}, which plays an important role at higher energy transfers and is directly relevant for the neutrino oscillation program.

\begin{acknowledgments}
Support by the Deutsche Forschungsgemeinschaft (DFG, German Research Foundation) – Project-ID 279384907 – SFB 1245, and through the Cluster of Excellence “Precision Physics, Fundamental Interactions, and Structure of Matter” (PRISMA+ EXC 2118/1, Project ID 390831469) is acknowledged.
This work used the DiRAC Data Intensive service (DIaL3) at the University of Leicester, managed by the University of Leicester Research Computing Service on behalf of the STFC DiRAC HPC Facility (www.dirac.ac.uk). The DiRAC service at Leicester was funded by BEIS, UKRI and STFC capital funding and STFC operations grants. DiRAC is part of the UKRI Digital Research Infrastructure.
Part of the calculations were also performed at the supercomputer Mogon at Johannes Gutenberg Universit\"at Mainz.
\end{acknowledgments}

\FloatBarrier

\bibliography{bib}

\end{document}